\documentclass[fleqn,10pt]{wlscirep}
\usepackage[utf8]{inputenc}
\usepackage[T1]{fontenc}
\usepackage{caption}
\usepackage{subcaption}
 \usepackage{multirow}
\title{A deep learning model for gastric diffuse-type adenocarcinoma classification in whole slide images}

\author[1]{Fahdi Kanavati}
\author[1,2,*]{Masayuki Tsuneki}

\affil[1]{Medmain Research, Medmain Inc., Fukuoka, 810-0042, Japan}
\affil[2]{Medmain Inc., Fukuoka, 810-0042, Japan}

\affil[*]{Corresponding author: tsuneki@medmain.com}

\begin{abstract}
Gastric diffuse-type adenocarcinoma represents a disproportionately high percentage of cases of gastric cancers occurring in the young, and its relative incidence seems to be on the rise. Usually it affects the body of the stomach, and presents shorter duration and worse prognosis compared with the differentiated (intestinal) type adenocarcinoma. The main difficulty encountered in the differential diagnosis of gastric adenocarcinomas occurs with the diffuse-type. As the cancer cells of diffuse-type adenocarcinoma are often single and inconspicuous in a background desmoplaia and inflammation, it can often be mistaken for a wide variety of non-neoplastic lesions including gastritis or reactive endothelial cells seen in granulation tissue.
In this study we trained deep learning models to classify gastric diffuse-type adenocarcinoma from WSIs. We evaluated the models on five test sets obtained from distinct sources, achieving receiver operator curve (ROC) area under the curves (AUCs) in the range of 0.95-0.99. The highly promising results demonstrate the potential of AI-based computational pathology for aiding pathologists in their diagnostic workflow system.

\end{abstract}
\begin{document}

\flushbottom
\maketitle

\thispagestyle{empty}

\section*{Introduction}

According to the global cancer statistics 2020\cite{Sung2021}, gastric cancer is amongst the most common leading causes of cancer related deaths in the world which is estimated 769,000 deaths and ranked fifth for incidence and fourth for mortality globally. Symptoms of gastric carcinoma tend to manifest only when it is at an advanced stage. The first sign is the detection of nodal, hepatic, and pulmonary metastases. In countries with a high incidence of gastric cancer, especially Japan, the increased use of endoscopic biopsy and cytology has resulted in the identification of early stage cases which has resulted in an increase in survival rates \cite{halvorsen1996diagnosis, iishi1986evaluation, Nagata1983, nashimoto2013gastric}. 
Microscopically, nearly all gastric carcinomas are of the adenocarcinoma (ADC) type and are composed of foveolar, mucopeptic, intestinal columnar, and goblet cell types \cite{Fiocca1987}. 
According to the Lauren classification \cite{LAURN1965} gastric ADCs are separated into intestinal and diffuse types. The intestinal-type shows well-defined glandular structures with papillae, tubules, or even solid areas. By contrast, the diffuse-type consists of poorly-differentiated type and signet ring cell carcinoma (SRCC). Diffuse-type ADC scatters and infiltrates widely, and its cells are small, uniform, and cohesive. Often these cells exhibit an SRCC appearance with the intracytoplasmic mucin pushing the nucleus of the neoplastic cells to the periphery. The amount of mucin present in these cells may be highly variable and difficult to appreciate in diffuse-type ADCs. 
Diffuse-type ADCs are more challenging to diagnose than other gastric carcinomas such as the intestinal-type. Diffuse-type cells are often single and inconspicuous in a background desmoplasia and inflammation, and they can often be mistaken for a variety of non-neoplastic lesions including gastritis or reactive endothelial cells in granulation tissues. Surgical pathologists are always on the lookout for signs of diffuse-type gastric adenocarcinoma when evaluating gastric biopsies.

Deep learning has found many successful applications in computational pathology in the past few years for tasks such as tumour and mutation classification, cell segmentation, and outcome prediction for a variety of organs and diseases \cite{yu2016predicting, hou2016patch, madabhushi2016image, litjens2016deep, kraus2016classifying, korbar2017deep, luo2017comprehensive, coudray2018classification, wei2019pathologist, gertych2019convolutional, bejnordi2017diagnostic, saltz2018spatial, campanella2019clinical, iizuka2020deep}. For stomach in particular,  Sharma et al. \cite{sharma2017deep} trained a model for carcinoma classification using a small training set of 11 WSIs, while Iizuka et al. \cite{iizuka2020deep} trained a deep learning model using a large dataset of 4,036 WSIs to classify gastric biopsy specimens into adenocarcinoma, adenoma, and non-neoplastic.

In this paper, we trained deep learning models for the classification of diffuse-type ADC in endoscopic biopsy specimen whole slide images (WSIs). To do so, we used two approaches: one-stage and two-stage. With the one-stage approach, the model was trained to directly classify diffuse-type ADC. With the two-stage approach, we used the model of Iizuka et al. \cite{iizuka2020deep} to first detect ADC, followed by a second stage model that subclassifies the detected ADC cases into diffuse-type ADC vs other ADC. For both approaches, we have used the partial transfer learning method\cite{kanavati2021partial} to fine-tune the models. We obtained models with ROC AUCs in the range in 0.95-0.99 for the five independent test sets, demonstrating the potential of such methods for aiding pathologists in their workflows.

\section*{Results}

\begin{figure}[ht]
\centering
\includegraphics[width=\linewidth]{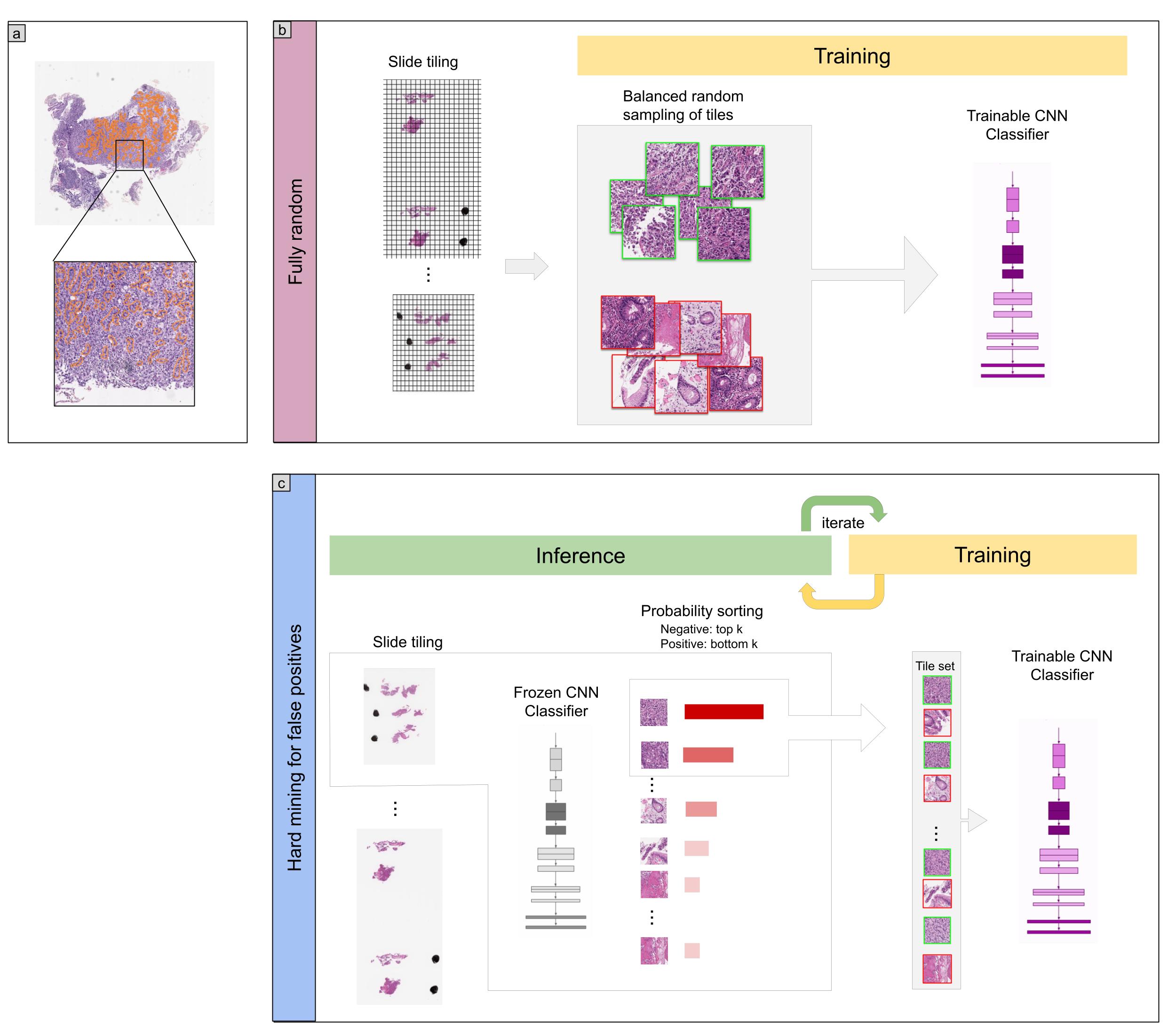}
\caption{Method overview. (a) An example of diffuse-type ADC annotation that was carried out digitally on the WSIs by pathologists. (b) The initial training consisted in fully-random balanced sampling of positive (diffuse-type ADC) and negative tiles to fine-tune the model. (b) Once there was no further improvement on the validation set after 2 epochs, the training switched into hard mining of tiles, which is an iterative process that alternates between training and inference. During the inference step, we applied the model in a sliding window fashion on all of the WSI and selected the k tiles with the highest probabilities if the WSI was negative, and k tiles with the lowest probabilities if the WSI was positive. The tiles were collected in a subset, and once the subset reached a given size, it was batched and used for training. This process allows training on the hardest examples and reduce false positives.}
\label{fig:overview}
\end{figure}

The aim of this study was to train a convolutional neural network (CNN) for the classification of diffuse-type ADC in biopsy WSIs. In order to apply a CNN on the large WSIs, we followed the commonly adopted approach of tiling the WSIs by extracting fixed-sized tiles over all the detected tissue regions (see methods section for more details). 
Overall, we trained four different models: (1) a two-stage method using existing model of Iizuka et al. \cite{iizuka2020deep} to first detect ADC, followed by a second model that detects diffuse-type ADC, both at x10 magnification; (2) a one-stage method for direct diffuse-type ADC classification at magnification x10 and a tile size of 224x224 px; (3) a one-stage method for direct diffuse-type ADC classification at magnification x20 and a tile size of 224x224 px; and (4) a one-stage method for direct diffuse-type ADC classification at magnification x20 and a tile size of 512x512 px. Figure \ref{fig:overview} provides an overview of the training of a given model.

\subsection*{Evaluation on five independent test sets from different sources}

\begin{figure}[ht]
\centering
 \begin{subfigure}[t]{0.47\textwidth}
         \centering
         \includegraphics[width=\textwidth]{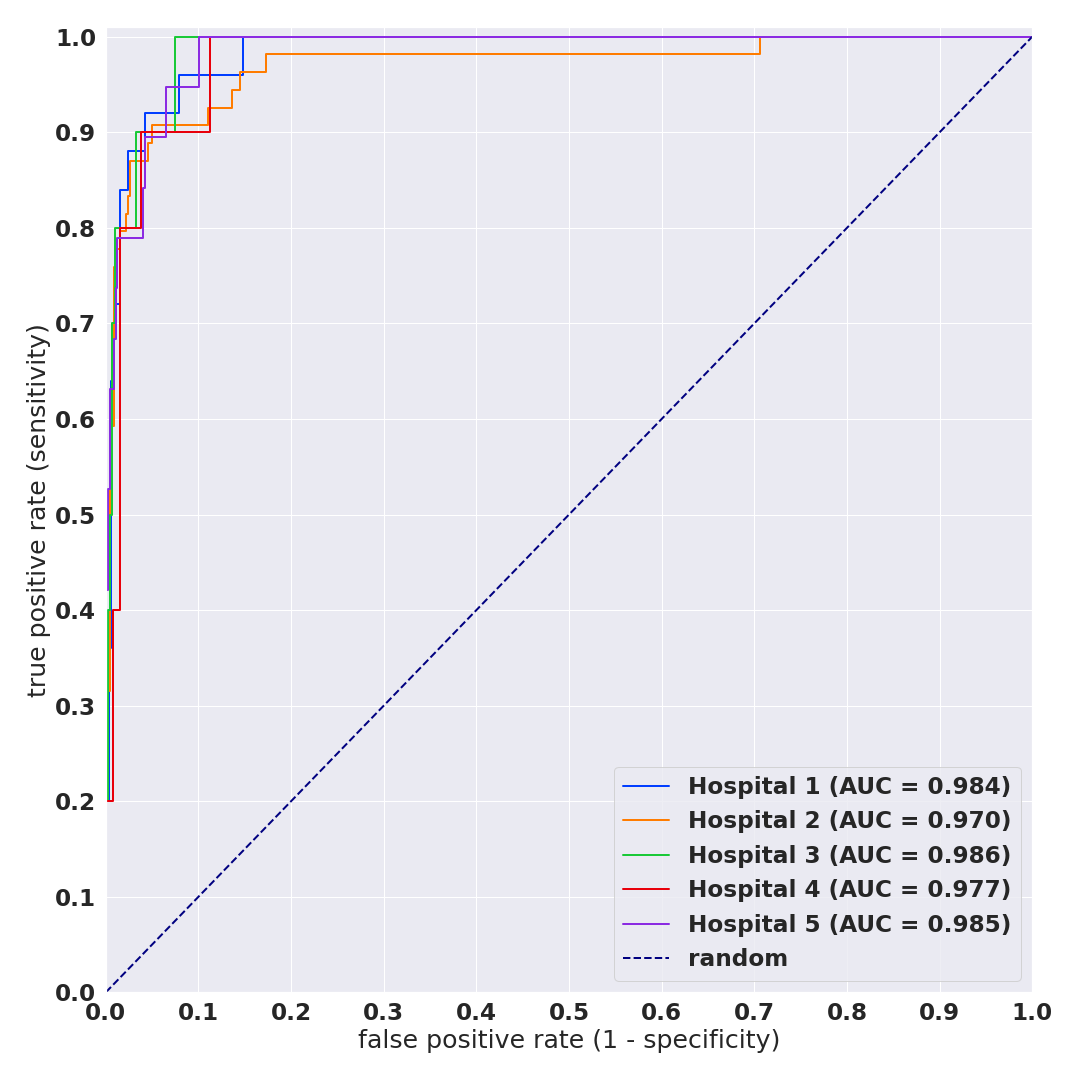}
         \caption{one-stage at magnification x20 and tile size 512x512px}
         \label{fig:1stage_x20_512}
     \end{subfigure}
     \hfill
     \begin{subfigure}[t]{0.47\textwidth}
         \centering
        \includegraphics[width=\linewidth]{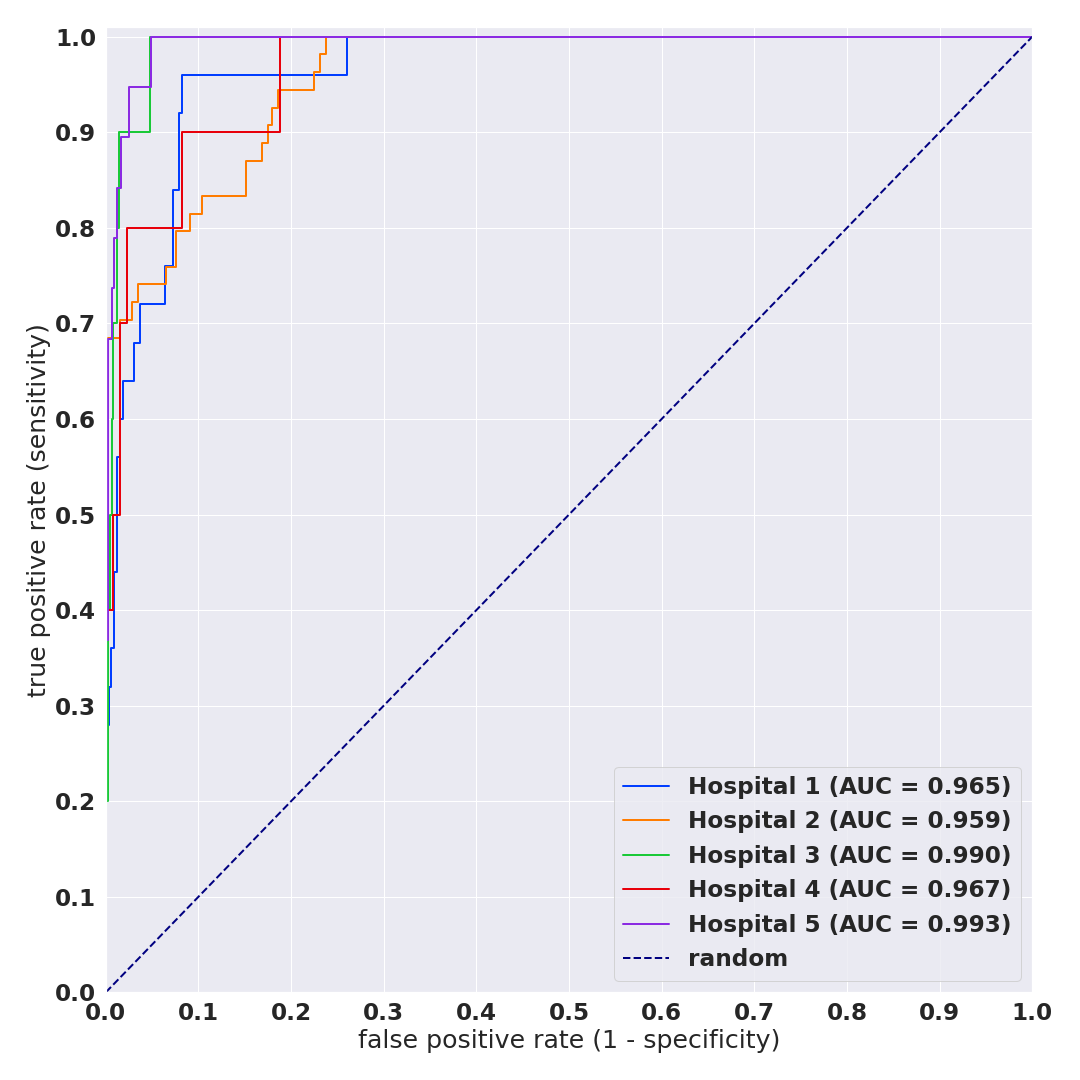}
\caption{two-stage at magnification x10 and tile size 224x224px}
         \label{fig:2stage_x10}
     \end{subfigure}
          \begin{subfigure}[t]{0.47\textwidth}
         \centering
        \includegraphics[width=\linewidth]{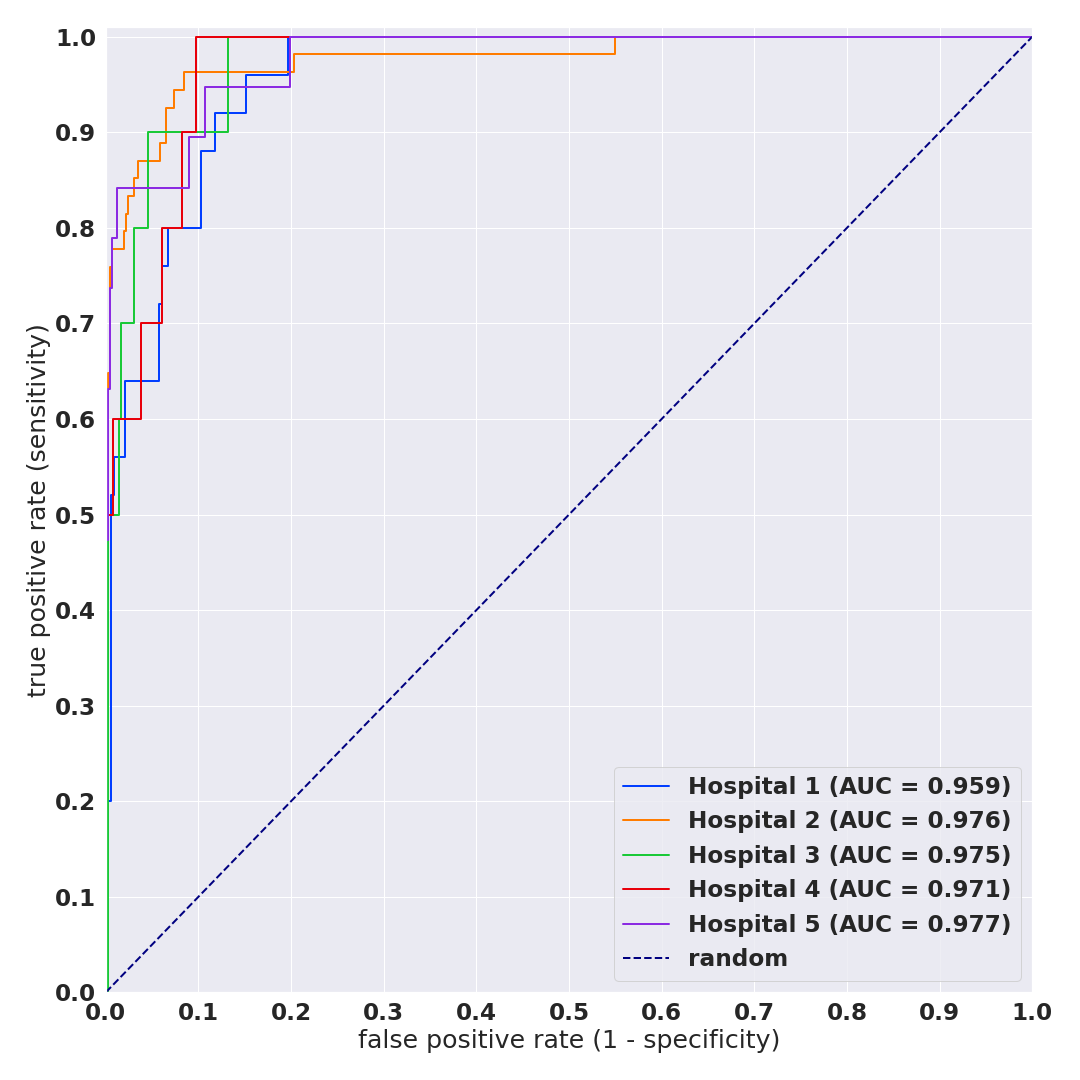}
\caption{one-stage at magnification x20 and tile size 224x224px}
         \label{fig:1stage_x20_224}
     \end{subfigure}
      \hfill
          \begin{subfigure}[t]{0.47\textwidth}
         \centering
        \includegraphics[width=\linewidth]{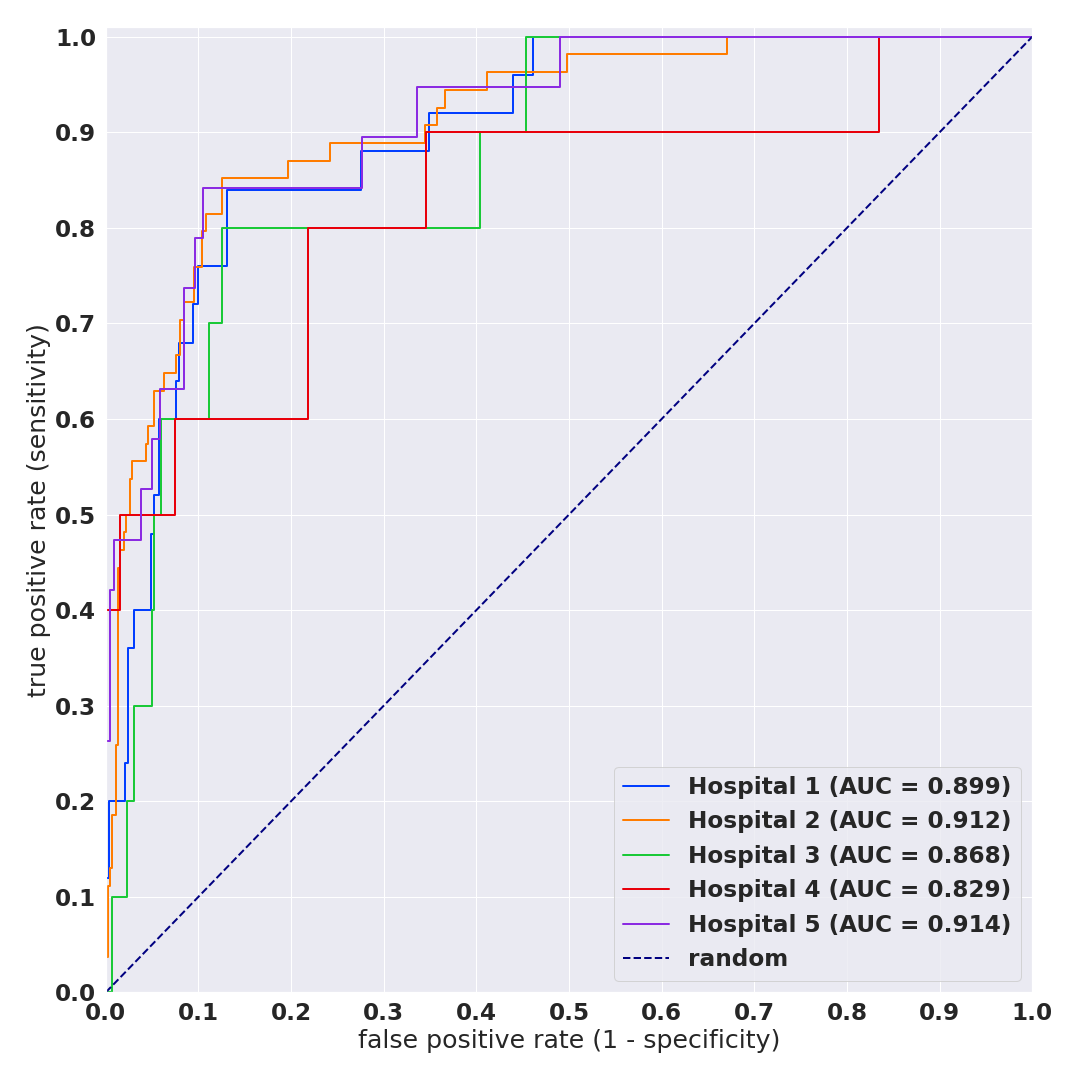}
\caption{one-stage at magnification x10 and tile size 224x224px.}
         \label{fig:1stage_x10_224}
     \end{subfigure}

\caption{\label{fig:roc_curves}ROC curves and corresponding AUCs for the test sets from five different hospitals using four different methods}
\end{figure}

\begin{figure}[ht]
\centering
\includegraphics[width=\linewidth]{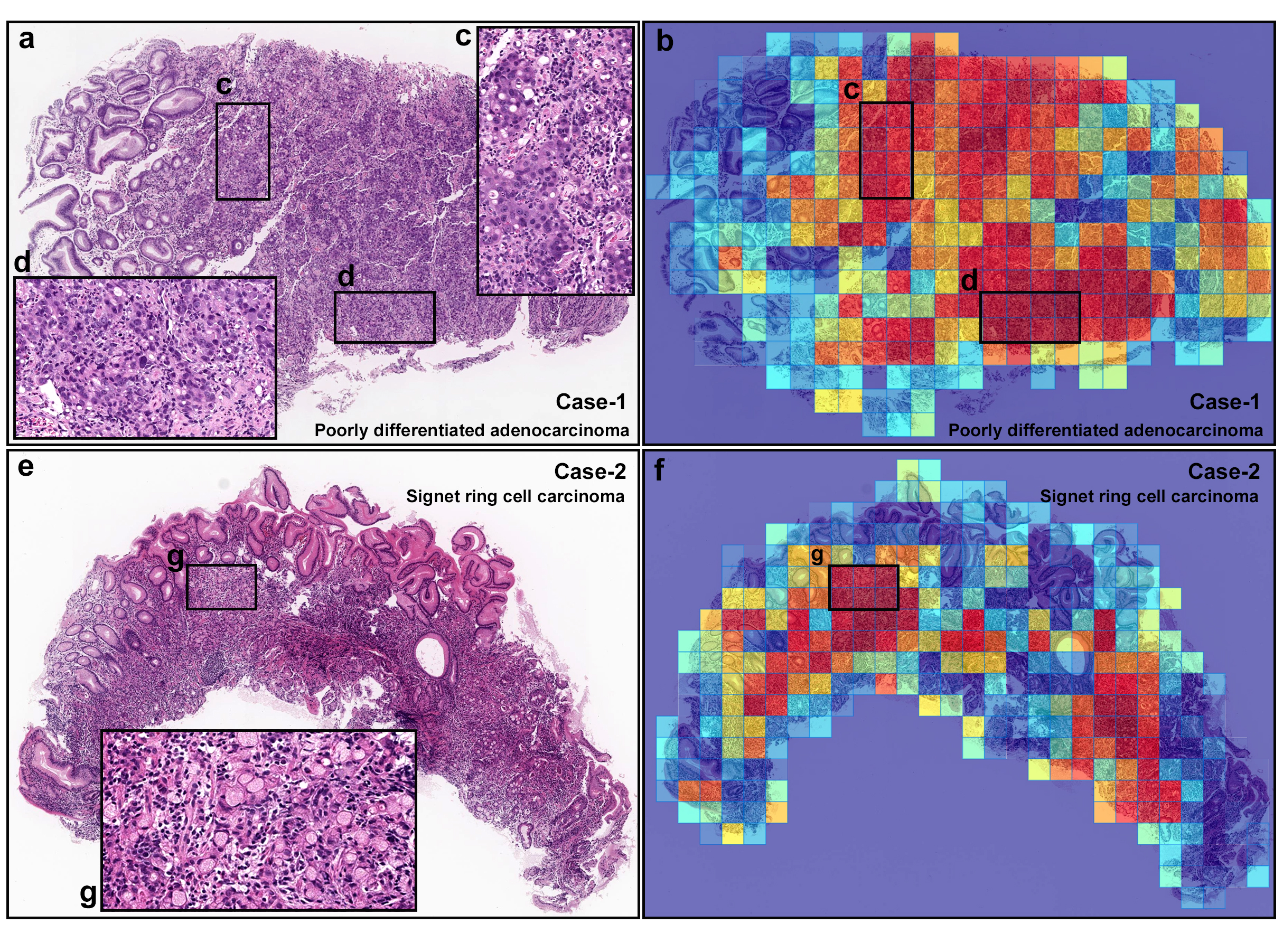}
\caption{Representative true positive diffuse-type gastric ADC cases from endoscopic biopsy test set. Case-1 (a-d) is a poorly-differentiated ADC and Case-2 (e-g) is a SRCC. Heatmap images (b and f) show true positive predictions of poorly-differentiated ADC (b) and SRCC (f) cells and they correspond respectively to (a) and (e) H\&E histopathology. The high magnification images (c, d, and g) show representative poorly-differentiated ADC (c and d) and SRCC (g) cellular morphology.}
\label{fig:tp}
\end{figure}

\begin{figure}[ht]
\centering
\includegraphics[width=\linewidth]{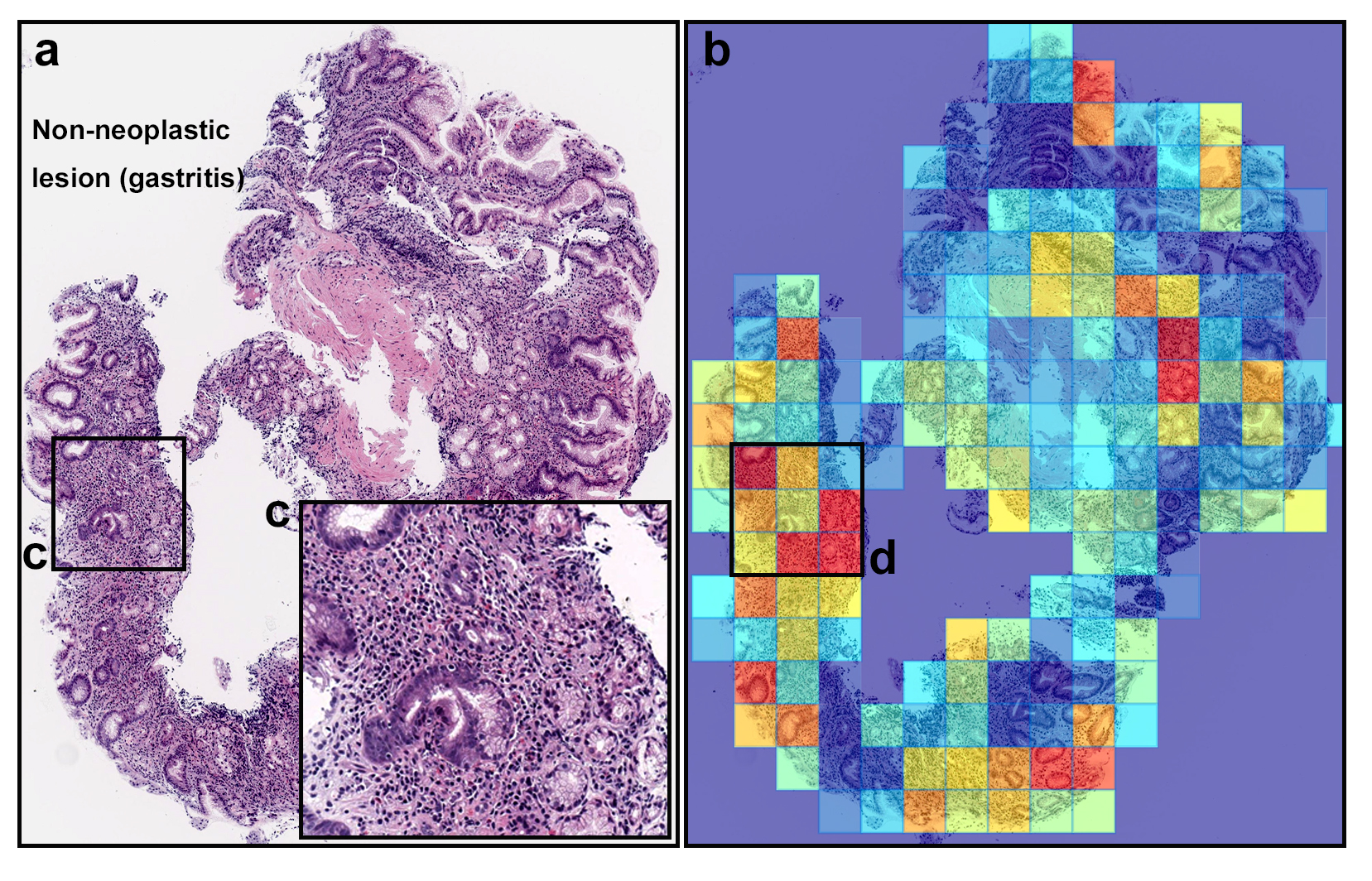}
\caption{A representative example of diffuse-type ADC false-positive prediction outputs. (a) is a non-neoplastic lesion (chronic gastritis). Heatmap images (b) exhibited false positive predictions of diffuse-type ADC. The inflammatory tissue with plasma cell infiltration (c) is the possible main cause of false positive (d) due to its analogous nuclear and cellular morphology to diffuse-type ADC cells.}
\label{fig:fp}
\end{figure}

\begin{figure}[ht]
\centering
\includegraphics[width=\linewidth]{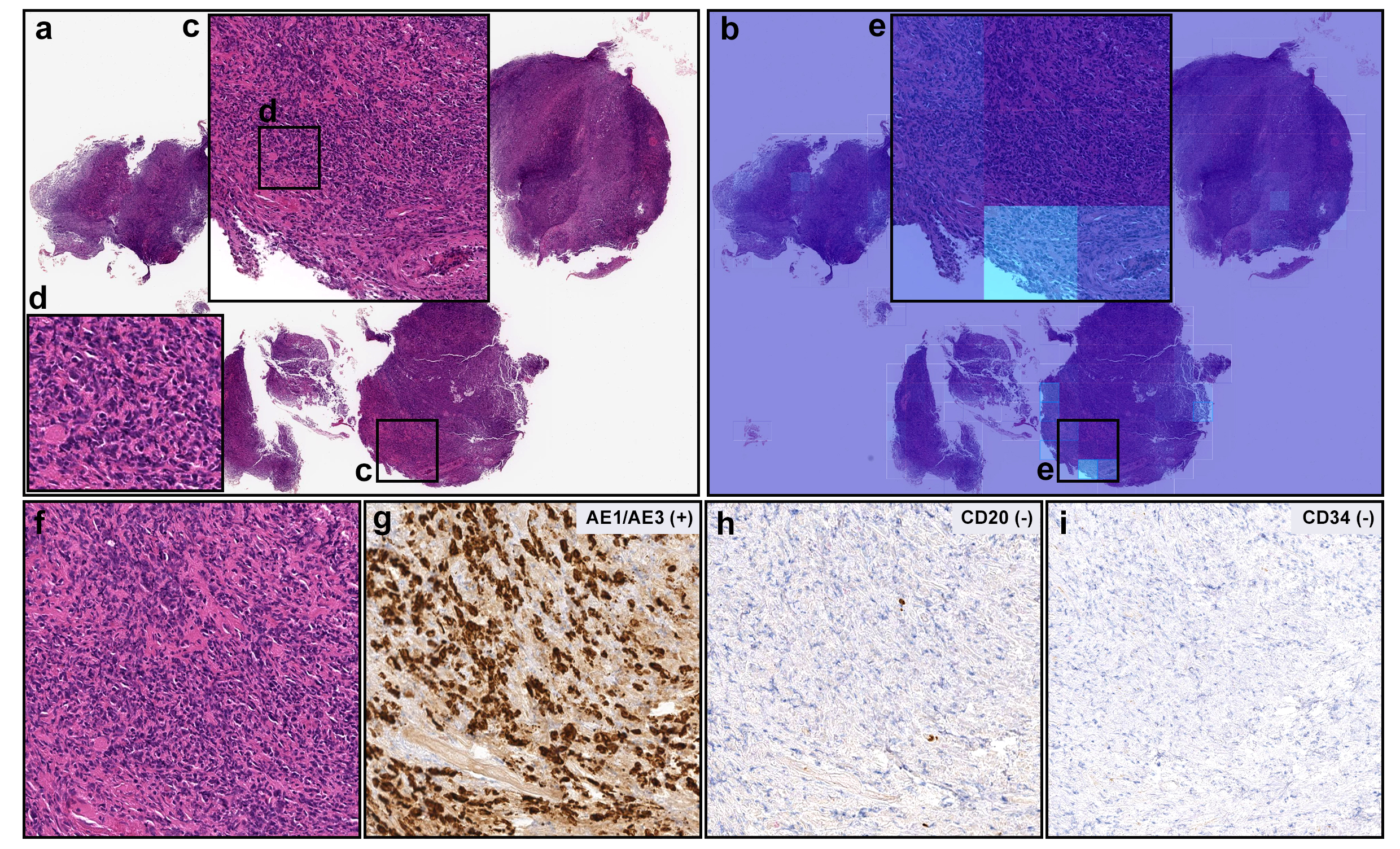}
\caption{A representative false negative case. In (a), there are numerous number of infiltrating degenerative cancer cells (c, d, f) which were not predicted as diffuse-type ADC cells on heatmap image (b, e) in necrotic and granulation tissues. After immunohistochemical stainings with AE1/AE3 (g), CD20 (h), and CD34 (i), infiltrating cancer cells (f) exhibited AE1/AE3 positive, CD20 negative, and CD34 negative, indicating cancer of epithelial origin (carcinoma). Therefore, histopathologically, this case was diagnosed as a diffuse-type differentiated ADC.}
\label{fig:fn}
\end{figure}

We evaluated our models on five test sets consisting of biopsy specimens originating each from a distinct hospital. Table \ref{tab:breakdown} breaks down the distribution of the WSIs in each test set. For each test set, we computed the ROC AUC for the WSI classification of diffuse-type ADC as well as the log loss, and we have summarised the results in Tab. \ref{tab:my_label} and Fig. \ref{fig:roc_curves}. Figures \ref{fig:tp}, \ref{fig:fp}, and \ref{fig:fn} show true positive, false positive, and false negative example heatmap outputs, respectively. 

\begin{table}[]
    \centering
\begin{tabular}{c|c|c |c}
\toprule
         method &      source &                  ROC AUC &                 Log loss \\
\midrule
 \multirow{5}{*}{two-stage x10 224} &  Hospital 1 &  0.9633 [0.9335, 0.9864] &  0.2892 [0.2251, 0.3396] \\
  &  Hospital 2 &   0.9590 [0.9365, 0.9771] &  0.1931 [0.1413, 0.2461] \\
  &  Hospital 3 &  0.9903 [0.9771, 0.9983] &  0.0651 [0.0435, 0.0948] \\
  &  Hospital 4 &  0.9669 [0.9209, 0.9972] &  0.3075 [0.2121, 0.4048] \\
  &  Hospital 5 &   0.9932 [0.9863, 0.999] &  0.1249 [0.0914, 0.1511] \\
 \hline
 \multirow{5}{*}{one-stage x20 512} &  Hospital 1 &  0.9835 [0.9776, 0.9973] &  0.6243 [0.5222, 0.7274] \\
  &  Hospital 2 &    0.9696 [0.9375, 0.990] &     0.3333 [0.2731, 0.400] \\
  &  Hospital 3 &   0.9862 [0.9824, 0.999] &   0.3850 [0.3294, 0.4484] \\
  &  Hospital 4 &  0.9774 [0.9369, 0.9968] &  1.1246 [0.9182, 1.3698] \\
  &  Hospital 5 &  0.9847 [0.9701, 0.9986] &    0.5789 [0.512, 0.662] \\
  \hline
 \multirow{5}{*}{one-stage x20 224} &  Hospital 1 &  0.9594 [0.9314, 0.9818] &   0.5874 [0.5324, 0.649] \\
  &  Hospital 2 &   0.9759 [0.9508, 0.993] &   0.1988 [0.1810, 0.2202] \\
  &  Hospital 3 &  0.9751 [0.9464, 0.9944] &  0.4177 [0.3942, 0.4501] \\
  &  Hospital 4 &  0.9714 [0.9281, 0.9935] &  0.9354 [0.8142, 1.0714] \\
  &  Hospital 5 &  0.9774 [0.9498, 0.9978] &  0.4428 [0.4128, 0.4682] \\
  \hline
 \multirow{5}{*}{one-stage x10 224} &  Hospital 1 &  0.8989 [0.8278, 0.9473] &  1.4383 [1.2866, 1.6029] \\
  &  Hospital 2 &  0.9118 [0.8712, 0.9476] &  0.2699 [0.2242, 0.3253] \\
  &  Hospital 3 &  0.8685 [0.7532, 0.9543] &  1.0267 [0.9322, 1.1403] \\
  &  Hospital 4 &  0.8293 [0.6485, 0.9601] &  1.8348 [1.5575, 2.1335] \\
  &  Hospital 5 &   0.9137 [0.8440, 0.9735] &  1.1132 [1.0001, 1.2146] \\
\bottomrule

\end{tabular}

    \caption{ROC AUC and log loss results for the four different models as evaluated on the five different test sets. First row group corresponds to the two-stage model, and the remaining three correspond to the one-stage models, with variations in magnification and tile size. Confidence intervals are between brackets.  \label{tab:my_label}}
   
\end{table}

\subsection*{Evaluation on surgical and frozen sections}

\begin{figure}[ht]
\centering
\includegraphics[width=0.8\linewidth]{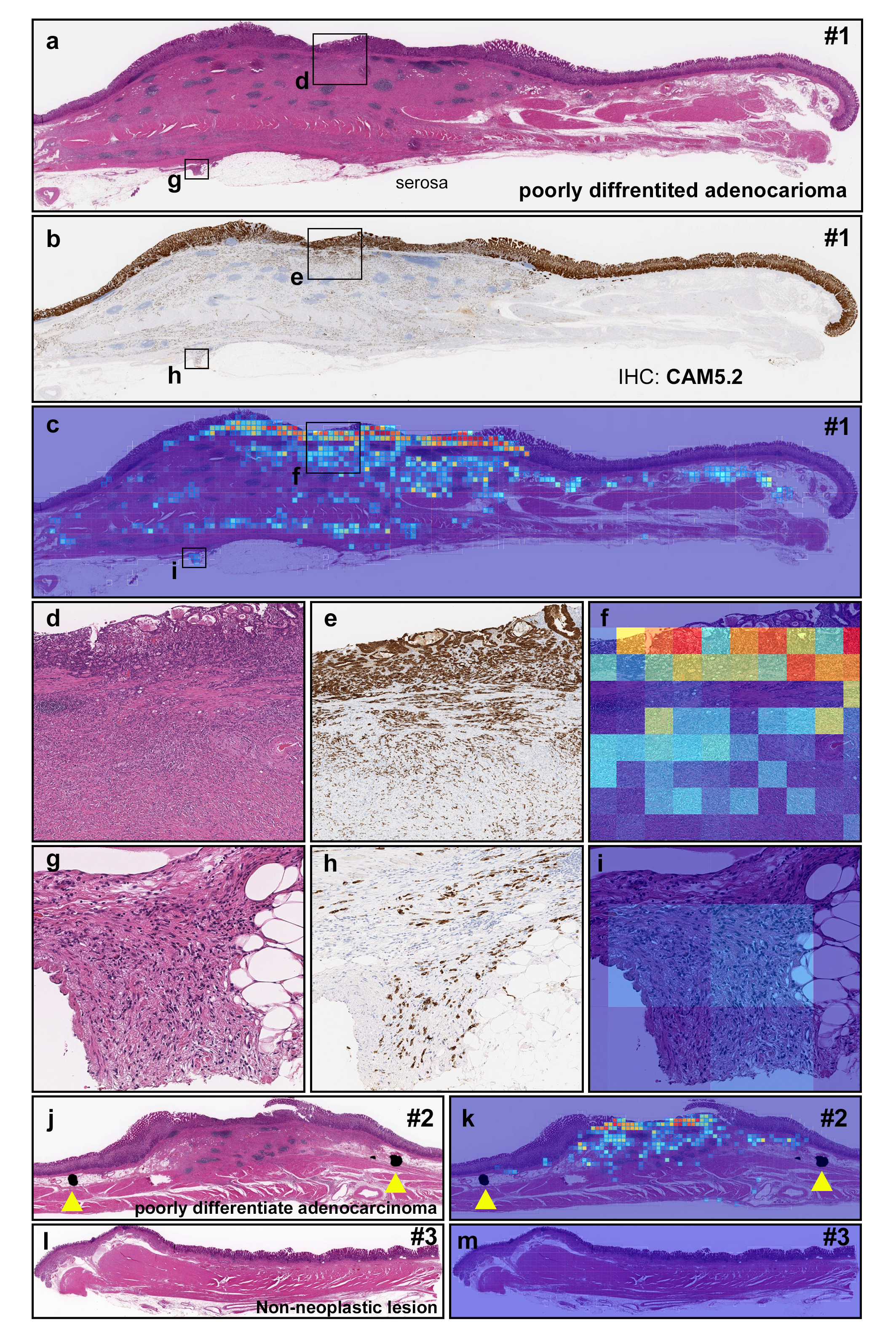}
\caption{A representative surgically resected case serial specimens for diffuse-type differentiated ADC. Serial specimens: \#1 (a-i), \#2 (j, k), and \#3 (l, m). In \#1 (a), diffuse-type differentiated ADC cells which are positive for CAM5.2 (b, e, f) invaded from submucosa (d, e) to subserosa (g, h). (c, f, i) show true positive probability heatmaps for invading diffuse-type differentiated ADC. In \#2 (j), true positive probability heatmap image for invading diffuse-type differentiated ADC (k) cell invading area was corresponded to surgical pathologists marked area with ink-dots (yellow-triangles) (j). \#3 (l) is a non-neoplastic tissue without any sign of cancer cell invading which is corresponded to the heatmap image (m).}
\label{fig:surgical}
\end{figure}

\begin{figure}[ht]
\centering
\includegraphics[width=\linewidth]{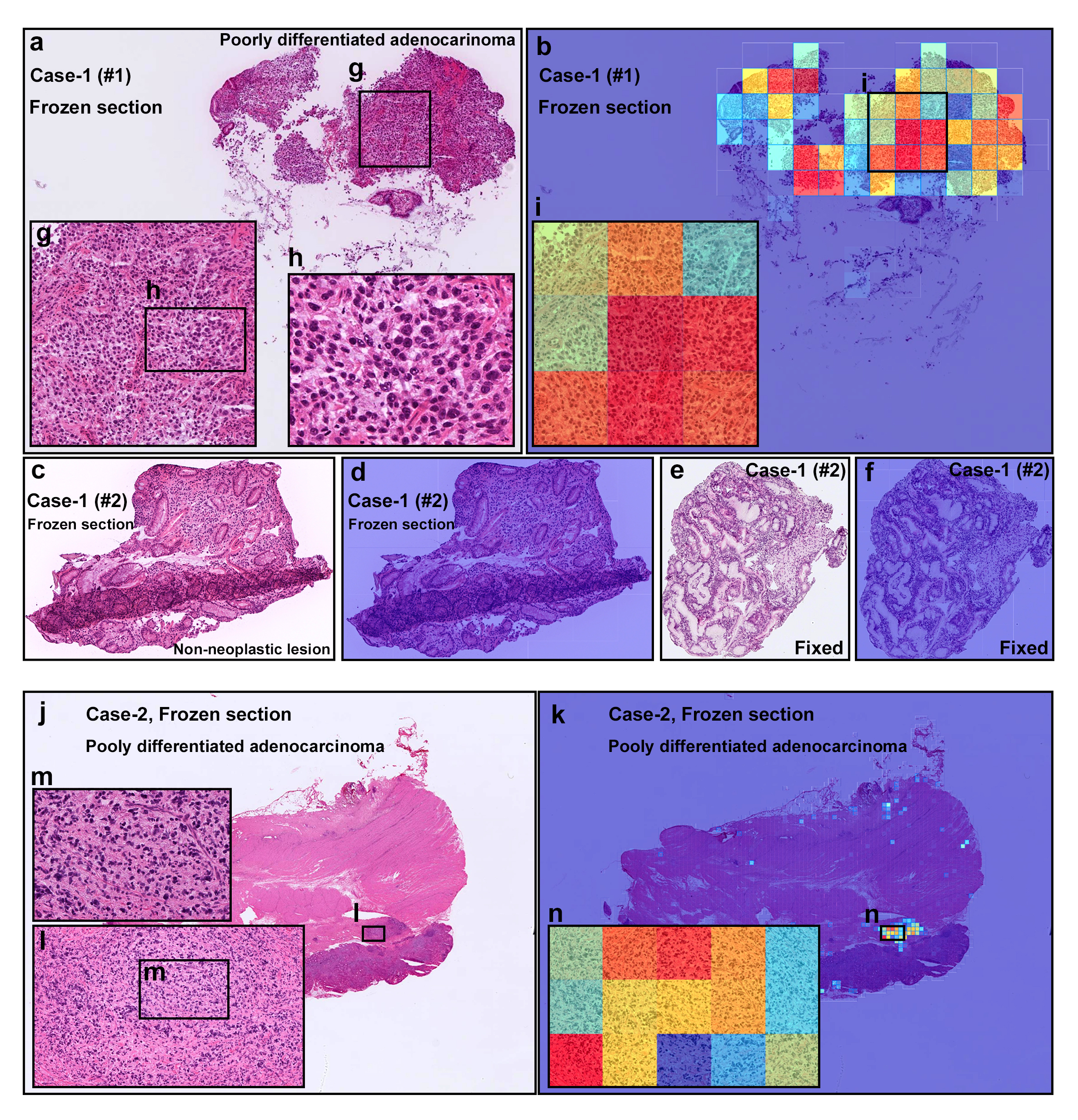}
\caption{Representative two cases of frozen section specimens for diffuse-type differentiated ADC. Case-1 consisted of two specimens (\#1 and \#2). In Case-1 (\#1) (a), the heatmap image (b) shows true positive predictions of diffuse-type differentiated ADC cells (g-i). In Case-1 (\#2) frozen section specimen (c), there was no cancer cells indicating non-neoplastic specimen which was corresponded to the heatmap image (d). The frozen section (\#2) was double-checked after conventional fixation (e). No cancer cells were observed in the fixed specimen as well (e) which was corresponded to the heatmap image (f). In Case-2 (j), the heatmap image (k) shows true positive predictions of diffuse-type differentiated ADC cells (l-n).}
\label{fig:frozen}
\end{figure}

In addition to the biopsy samples, we have applied the model on the small number of surgical and frozen sections. Figures \ref{fig:surgical} and \ref{fig:frozen} show example output predictions on such cases. We see the model was capable of detection diffuse-type ADC on such sections.

\section*{Discussion}
In this work, we trained models for the classification of gastric diffuse-type ADC from biopsy WSIs.
We used the partial transfer learning approach with a hard mining of false positives to train the models on a dataset obtained from a single hospital, and we evaluated them on five different test sets originating from different hospitals. Overall, we obtained high ROC AUCs in the range of 0.95-0.99.

The best performing models were the one-stage model at x20 magnification and 512x512px tile size and the 2-stage model at x10 magnification and 224x224px tile size. For the one-stage model, training at x20 magnification led to an increase in performance, where the average ROC AUC increased from 0.87 to 0.97 for the five test sets. The increase in magnification was most likely essential in decreasing the false positive rate. Despite being at x10 magnification, the two-stage model still performed well potentially due to having been trained on a much larger datasets (n=4,036) and the use of the RNN model which aims at reducing the false-positives.

The trained model was able to detect well both poorly-differentiated ADC and SRCC cells (see Fig. \ref{fig:tp} for an example representative case). 
The majority of false positives occurred on gastritis cases due to the similarity between diffuse-type ADC and inflammatory cells especially plasma cells (see Fig. \ref{fig:fp}).

Diffuse-type gastric ADCs composed are composed of diffuse-type cohesive carcinoma and SRCCs \cite{hu2012gastric}, and they show an aggressive biological behavior and poor prognosis\cite{Lee2017}. In a previous report, patients with SRCC and diffuse-type differentiated ADC in advanced stages demonstrated significantly lower 10-year overall survival rates than the survival rates of patients with advanced differentiated-type ADCs\cite{Chon2017}. The availability of a tool that can aid pathologists in the diagnosis of diffuse-type ADC could potentially accelerate their diagnostic workflow.

\section*{Methods}

\subsection*{Clinical cases and pathological records}

For the present retrospective study, a total of 2,929 endoscopic biopsy cases of human gastric epithelial lesions HE (hematoxylin \& eosin) stained histopathological specimens were collected from the surgical pathology files of five hospitals: International University of Health and Welfare, Mita Hospital (Tokyo), Kamachi Group Hospitals (Fukuoka), Haradoi Hospital (Fukuoka), and Nishi-Fukuoka Hospital (Fukuoka) after histopathological review of those specimens by surgical pathologists.
The experimental protocol was approved by the ethical board of the International University of Health and Welfare (No. 19-Im-007), Kamachi Group Hospitals, Haradoi Hospital, and Nishi-Fukuoka Hospital. All research activities complied with all relevant ethical regulations and were performed in accordance with relevant guidelines and regulations in the all hospitals mentioned above. Informed consent to use histopathological samples and pathological diagnostic reports for research purposes had previously been obtained from all patients prior to the surgical procedures at all hospitals, and the opportunity for refusal to participate in research had been guaranteed by an opt-out manner. The test cases were selected randomly, so the obtained ratios reflected a real clinical scenario as much as possible. All WSIs were scanned at a magnification of x20.

\subsection*{Dataset and annotations}

The pathologists excluded cases that were inappropriate or of poor quality for this study. The diagnosis of each WSI was verified by at least two pathologists. Table \ref{tab:breakdown} breaks down the distribution of the datasets into training, validation, and test sets. Hospitals which provided histopathological cases were anonymised (e.g., Hospital 1-5). The training and test sets were solely composed of WSIs of endoscopic biopsy specimens. The patients' pathological records were used to extract the WSIs' pathological diagnoses. 353 WSIs from the training and validation sets had a diffuse-type ADC diagnosis. They were manually annotated by a group of two surgical pathologists who perform routine histopathological diagnoses. The pathologists carried out detailed cellular-level annotations by free-hand drawing around diffuse-type ADC cells that corresponded to poorly-differentiated ADC or SRCC. The other ADC (n=571) and non-neoplastic subsets (n=1,116) of the training and validation sets were not annotated and the entire tissue areas within the WSIs were used. Each annotated WSI was observed by at least two pathologists, with the final checking and verification performed by a senior pathologist.

\begin{table}[ht]
\centering
\begin{tabular}{l|c||ccc||c}
 Set                      &  Source           & diffuse-type ADC & other ADC & non-neoplastic & Total \\
\hline
2-stage: train                & Hospital 2          & 333        & 541       & -       & 874 \\
2-stage: validation & Hospital 2 & 20  & 30 & - & 50 \\ \hline 
1-stage: train                 &     Hospital 2      &  333        &  541       & 1076      & 1950 \\
1-stage: validation & Hospital 2 & 20  & 30 & 40 & 90 \\
                      \hline
\multirow{5}{*}{Test} & Hospital 1 & 25         & 96        & 234     & 355   \\
                      & Hospital 2 & 54         & 55        & 407     & 516   \\
                      & Hospital 3 & 10         & 12        & 473     & 495   \\
                      & Hospital 4 & 10         & 20        & 113     & 143   \\
                      & Hospital 5 & 19         & 38        & 439     & 496   \\
                      \hline
\end{tabular}

\caption{\label{tab:breakdown}Distribution of WSIs in the training, validation, and test sets.}
\end{table}

\subsection*{Deep learning models}

For the detection of diffuse-type ADC, we used two approaches: one-stage and two-stage.
The one-stage approach consisted in training the CNN as a binary classifier to directly classify diffuse-type ADC.
The two-stage approach consisted in combining the output from an existing model\cite{iizuka2020deep} that differentiates between ADC, adenoma, and non-neoplastic \cite{iizuka2020deep}, followed by a model trained to differentiate between diffuse-type ADC and other ADC. 
We trained all the models using the partial fine-tuning approach \cite{kanavati2021partial}. This method simply consists in using the weights of an existing pre-trained model and only fine-tuning the affine parameters of the batch normalisation layers and the final classification layer. We have used the EfficientNetB1\cite{tan2019efficientnet} model starting with pre-trained weights on ImageNet. The total number of trainable parameters was only 63,329.

To apply the CNN on the WSIs, we performed slide tiling by extracting square tiles from tissue regions. On a given WSI, we detected the tissue regions and eliminated most of the white background by performing a thresholding on a grayscale version of the WSI using Otsu's method \cite{otsu1979threshold}. During prediction, we perform the tiling in a sliding window fashion, using a fixed-size stride, to obtain predictions for all the tissue regions. During training, we initially performed random balanced sampling of tiles from the tissue regions, where we tried to maintain an equal balance of each label in the training batch. To do so, we placed the WSIs in a shuffled queue such that we looped over the labels in succession (i.e. we alternated between picking a WSI with a positive label and a negative label). Once a WSI was selected, we randomly sampled $\frac{\text{batch size}}{\text{num labels}}$ tiles from each WSI to form a balanced batch. To maintain the balance on the WSI, we over-sampled from the WSIs to ensure the model trains on tiles from all of the WSIs in each epoch. We then switched into hard mining of tiles once there was no longer any improvement on the validation set after two epochs. To perform the hard mining, we alternated between training and inference. During inference, the CNN was applied in a sliding window fashion on all of the tissue regions in the WSI, and we then selected the $k$ tiles with the highest probability for being positive if the WSI was negative and the $k$ tiles with the lowest probability for being positive if the WSI was positive. This step effectively selects the hard examples which the model is struggling with. The selected tiles were placed in a training subset, and once that subset contained $N$ tiles, the training was run. This method is similar to the weakly supervised training method as described by Kanavati et al.\cite{kanavati2020weakly}. We used $k = 16$, $N=256$, and a batch size of $32$.

From the WSIs with diffuse-type ADC, we sampled tiles based on the free-hand annotations. If the WSI contained annotations for cancer cells, then we only sampled tiles from the annotated regions as follows: if the annotation was smaller than the tile size, then we sampled the tile at the centre of the annotation regions; otherwise, if the annotation was larger than the tile size, then we subdivided the annotated regions into overlapping grids and sampled tiles. Most of the annotations were smaller than the tile size. On the other hand, if the WSI did not contain diffuse-type ADC, then we freely sampled from the entire tissue regions.

The first stage model \cite{iizuka2020deep} is based on the InceptionV3 architecture\cite{szegedy2016rethinking} is followed by a single layer recurrent neural network. It was trained with an input tile size of $512\times512$ px on WSIs with a magnification of x10. 
As the 2nd stage model was only trained on ADC, we used the product of the probability outputs to compute the probability that a given WSI has diffuse-type ADC:

$$P(\text{diffuse-type ADC}) = P_2(\text{diffuse-type ADC}| \text{ADC}) \times P_1(\text{ADC}),$$ where $P_1(\text{ADC})$ is the probability output from the 1st stage model and $P_2(\text{diffuse-type ADC}| \text{ADC})$ is the probability from the 2nd stage model.

We trained the models with the Adam optimisation algorithm \cite{kingma2014adam} with the following parameters: $beta_1=0.9$, $beta_2=0.999$, and a batch size of 32. We used a starting learning rate of $0.001$ when training the model from scratch, and $0.0001$ when fine-tuning. We applied a learning rate decay of $0.95$ every $2$ epochs. We used the categorical cross entropy loss function. We used early stopping by tracking the performance of the model on a validation set, and training was stopped automatically when there was no further improvement on the validation loss for 10 epochs. The model with the lowest validation loss was chosen as the final model.

\subsection*{Software and statistical analysis}

We implemented the models using TensorFlow\cite{tensorflow2015-whitepaper}. We calculated the AUCs in python using the scikit-learn package\cite{scikit-learn} and performed the plotting using matplotlib\cite{Hunter:2007}. We performed image processing, such as the thresholding with scikit-image \cite{scikit-image}. We computed the 95\% CIs estimates using the bootstrap method\cite{efron1994introduction} with 1000 iterations. We used openslide\cite{goode2013openslide} to perform real-time slide tiling.

\subsection*{Data availability}

Due to specific institutional requirements governing privacy protection, datasets used in this study are not publicly available.

\bibliography{main}

\begin{thebibliography}{10}
\urlstyle{rm}
\expandafter\ifx\csname url\endcsname\relax
  \def\url#1{\texttt{#1}}\fi
\expandafter\ifx\csname urlprefix\endcsname\relax\def\urlprefix{URL }\fi
\expandafter\ifx\csname doiprefix\endcsname\relax\def\doiprefix{DOI: }\fi
\providecommand{\bibinfo}[2]{#2}
\providecommand{\eprint}[2][]{\url{#2}}

\bibitem{Sung2021}
\bibinfo{author}{Sung, H.} \emph{et~al.}
\newblock \bibinfo{journal}{\bibinfo{title}{Global cancer statistics 2020:
  {GLOBOCAN} estimates of incidence and mortality worldwide for 36 cancers in
  185 countries}}.
\newblock {\emph{\JournalTitle{{CA}: A Cancer Journal for Clinicians}}}
  \doiprefix\url{10.3322/caac.21660} (\bibinfo{year}{2021}).

\bibitem{halvorsen1996diagnosis}
\bibinfo{author}{Halvorsen~Jr, R.~A.}, \bibinfo{author}{Yee, J.} \&
  \bibinfo{author}{McCormick, V.~D.}
\newblock \bibinfo{title}{Diagnosis and staging of gastric cancer.}
\newblock In \emph{\bibinfo{booktitle}{Seminars in oncology}},
  vol.~\bibinfo{volume}{23}, \bibinfo{pages}{325--335} (\bibinfo{year}{1996}).

\bibitem{iishi1986evaluation}
\bibinfo{author}{Iishi, H.}, \bibinfo{author}{Yamamoto, R.},
  \bibinfo{author}{Tatsuta, M.} \& \bibinfo{author}{Okuda, S.}
\newblock \bibinfo{journal}{\bibinfo{title}{Evaluation of fine-needle
  aspiration biopsy under direct vision gastrofiberscopy in diagnosis of
  diffusely infiltrative carcinoma of the stomach}}.
\newblock {\emph{\JournalTitle{Cancer}}} \textbf{\bibinfo{volume}{57}},
  \bibinfo{pages}{1365--1369} (\bibinfo{year}{1986}).

\bibitem{Nagata1983}
\bibinfo{author}{Nagata, T.}, \bibinfo{author}{Ikeda, M.} \&
  \bibinfo{author}{Nakayama, F.}
\newblock \bibinfo{journal}{\bibinfo{title}{Changing state of gastric cancer in
  japan}}.
\newblock {\emph{\JournalTitle{The American Journal of Surgery}}}
  \textbf{\bibinfo{volume}{145}}, \bibinfo{pages}{226--233},
  \doiprefix\url{10.1016/0002-9610(83)90068-5} (\bibinfo{year}{1983}).

\bibitem{nashimoto2013gastric}
\bibinfo{author}{Nashimoto, A.} \emph{et~al.}
\newblock \bibinfo{journal}{\bibinfo{title}{Gastric cancer treated in 2002 in
  japan: 2009 annual report of the jgca nationwide registry}}.
\newblock {\emph{\JournalTitle{Gastric cancer}}} \textbf{\bibinfo{volume}{16}},
  \bibinfo{pages}{1--27} (\bibinfo{year}{2013}).

\bibitem{Fiocca1987}
\bibinfo{author}{Fiocca, R.} \emph{et~al.}
\newblock \bibinfo{journal}{\bibinfo{title}{Characterization of four main cell
  types in gastric cancer: Foveolar, mucopeptic, intestinal columnar and goblet
  cells}}.
\newblock {\emph{\JournalTitle{Pathology - Research and Practice}}}
  \textbf{\bibinfo{volume}{182}}, \bibinfo{pages}{308--325},
  \doiprefix\url{10.1016/s0344-0338(87)80066-3} (\bibinfo{year}{1987}).

\bibitem{LAURN1965}
\bibinfo{author}{LAUR{\'{E}}N, P.}
\newblock \bibinfo{journal}{\bibinfo{title}{{THE} {TWO} {HISTOLOGICAL} {MAIN}
  {TYPES} {OF} {GASTRIC} {CARCINOMA}: {DIFFUSE} {AND} {SO}-{CALLED}
  {INTESTINAL}-{TYPE} {CARCINOMA}}}.
\newblock {\emph{\JournalTitle{Acta Pathologica Microbiologica Scandinavica}}}
  \textbf{\bibinfo{volume}{64}}, \bibinfo{pages}{31--49},
  \doiprefix\url{10.1111/apm.1965.64.1.31} (\bibinfo{year}{1965}).

\bibitem{yu2016predicting}
\bibinfo{author}{Yu, K.-H.} \emph{et~al.}
\newblock \bibinfo{journal}{\bibinfo{title}{Predicting non-small cell lung
  cancer prognosis by fully automated microscopic pathology image features}}.
\newblock {\emph{\JournalTitle{Nature communications}}}
  \textbf{\bibinfo{volume}{7}}, \bibinfo{pages}{12474} (\bibinfo{year}{2016}).

\bibitem{hou2016patch}
\bibinfo{author}{Hou, L.} \emph{et~al.}
\newblock \bibinfo{title}{Patch-based convolutional neural network for whole
  slide tissue image classification}.
\newblock In \emph{\bibinfo{booktitle}{Proceedings of the IEEE Conference on
  Computer Vision and Pattern Recognition}}, \bibinfo{pages}{2424--2433}
  (\bibinfo{year}{2016}).

\bibitem{madabhushi2016image}
\bibinfo{author}{Madabhushi, A.} \& \bibinfo{author}{Lee, G.}
\newblock \bibinfo{journal}{\bibinfo{title}{Image analysis and machine learning
  in digital pathology: Challenges and opportunities}}.
\newblock {\emph{\JournalTitle{Medical Image Analysis}}}
  \textbf{\bibinfo{volume}{33}}, \bibinfo{pages}{170--175}
  (\bibinfo{year}{2016}).

\bibitem{litjens2016deep}
\bibinfo{author}{Litjens, G.} \emph{et~al.}
\newblock \bibinfo{journal}{\bibinfo{title}{Deep learning as a tool for
  increased accuracy and efficiency of histopathological diagnosis}}.
\newblock {\emph{\JournalTitle{Scientific reports}}}
  \textbf{\bibinfo{volume}{6}}, \bibinfo{pages}{26286} (\bibinfo{year}{2016}).

\bibitem{kraus2016classifying}
\bibinfo{author}{Kraus, O.~Z.}, \bibinfo{author}{Ba, J.~L.} \&
  \bibinfo{author}{Frey, B.~J.}
\newblock \bibinfo{journal}{\bibinfo{title}{Classifying and segmenting
  microscopy images with deep multiple instance learning}}.
\newblock {\emph{\JournalTitle{Bioinformatics}}} \textbf{\bibinfo{volume}{32}},
  \bibinfo{pages}{i52--i59} (\bibinfo{year}{2016}).

\bibitem{korbar2017deep}
\bibinfo{author}{Korbar, B.} \emph{et~al.}
\newblock \bibinfo{journal}{\bibinfo{title}{Deep learning for classification of
  colorectal polyps on whole-slide images}}.
\newblock {\emph{\JournalTitle{Journal of pathology informatics}}}
  \textbf{\bibinfo{volume}{8}} (\bibinfo{year}{2017}).

\bibitem{luo2017comprehensive}
\bibinfo{author}{Luo, X.} \emph{et~al.}
\newblock \bibinfo{journal}{\bibinfo{title}{Comprehensive computational
  pathological image analysis predicts lung cancer prognosis}}.
\newblock {\emph{\JournalTitle{Journal of Thoracic Oncology}}}
  \textbf{\bibinfo{volume}{12}}, \bibinfo{pages}{501--509}
  (\bibinfo{year}{2017}).

\bibitem{coudray2018classification}
\bibinfo{author}{Coudray, N.} \emph{et~al.}
\newblock \bibinfo{journal}{\bibinfo{title}{Classification and mutation
  prediction from non--small cell lung cancer histopathology images using deep
  learning}}.
\newblock {\emph{\JournalTitle{Nature medicine}}}
  \textbf{\bibinfo{volume}{24}}, \bibinfo{pages}{1559--1567}
  (\bibinfo{year}{2018}).

\bibitem{wei2019pathologist}
\bibinfo{author}{Wei, J.~W.} \emph{et~al.}
\newblock \bibinfo{journal}{\bibinfo{title}{Pathologist-level classification of
  histologic patterns on resected lung adenocarcinoma slides with deep neural
  networks}}.
\newblock {\emph{\JournalTitle{Scientific reports}}}
  \textbf{\bibinfo{volume}{9}}, \bibinfo{pages}{1--8} (\bibinfo{year}{2019}).

\bibitem{gertych2019convolutional}
\bibinfo{author}{Gertych, A.} \emph{et~al.}
\newblock \bibinfo{journal}{\bibinfo{title}{Convolutional neural networks can
  accurately distinguish four histologic growth patterns of lung adenocarcinoma
  in digital slides}}.
\newblock {\emph{\JournalTitle{Scientific reports}}}
  \textbf{\bibinfo{volume}{9}}, \bibinfo{pages}{1483} (\bibinfo{year}{2019}).

\bibitem{bejnordi2017diagnostic}
\bibinfo{author}{Bejnordi, B.~E.} \emph{et~al.}
\newblock \bibinfo{journal}{\bibinfo{title}{Diagnostic assessment of deep
  learning algorithms for detection of lymph node metastases in women with
  breast cancer}}.
\newblock {\emph{\JournalTitle{Jama}}} \textbf{\bibinfo{volume}{318}},
  \bibinfo{pages}{2199--2210} (\bibinfo{year}{2017}).

\bibitem{saltz2018spatial}
\bibinfo{author}{Saltz, J.} \emph{et~al.}
\newblock \bibinfo{journal}{\bibinfo{title}{Spatial organization and molecular
  correlation of tumor-infiltrating lymphocytes using deep learning on
  pathology images}}.
\newblock {\emph{\JournalTitle{Cell reports}}} \textbf{\bibinfo{volume}{23}},
  \bibinfo{pages}{181--193} (\bibinfo{year}{2018}).

\bibitem{campanella2019clinical}
\bibinfo{author}{Campanella, G.} \emph{et~al.}
\newblock \bibinfo{journal}{\bibinfo{title}{Clinical-grade computational
  pathology using weakly supervised deep learning on whole slide images}}.
\newblock {\emph{\JournalTitle{Nature medicine}}}
  \textbf{\bibinfo{volume}{25}}, \bibinfo{pages}{1301--1309}
  (\bibinfo{year}{2019}).

\bibitem{iizuka2020deep}
\bibinfo{author}{Iizuka, O.} \emph{et~al.}
\newblock \bibinfo{journal}{\bibinfo{title}{Deep learning models for
  histopathological classification of gastric and colonic epithelial tumours}}.
\newblock {\emph{\JournalTitle{Scientific reports}}}
  \textbf{\bibinfo{volume}{10}}, \bibinfo{pages}{1--11} (\bibinfo{year}{2020}).

\bibitem{sharma2017deep}
\bibinfo{author}{Sharma, H.}, \bibinfo{author}{Zerbe, N.},
  \bibinfo{author}{Klempert, I.}, \bibinfo{author}{Hellwich, O.} \&
  \bibinfo{author}{Hufnagl, P.}
\newblock \bibinfo{journal}{\bibinfo{title}{Deep convolutional neural networks
  for automatic classification of gastric carcinoma using whole slide images in
  digital histopathology}}.
\newblock {\emph{\JournalTitle{Computerized Medical Imaging and Graphics}}}
  \textbf{\bibinfo{volume}{61}}, \bibinfo{pages}{2--13} (\bibinfo{year}{2017}).

\bibitem{kanavati2021partial}
\bibinfo{author}{Kanavati, F.} \& \bibinfo{author}{Tsuneki, M.}
\newblock \bibinfo{journal}{\bibinfo{title}{Partial transfusion: on the
  expressive influence of trainable batch norm parameters for transfer
  learning}}.
\newblock {\emph{\JournalTitle{arXiv preprint arXiv:2102.05543}}}
  (\bibinfo{year}{2021}).

\bibitem{hu2012gastric}
\bibinfo{author}{Hu, B.} \emph{et~al.}
\newblock \bibinfo{journal}{\bibinfo{title}{Gastric cancer: Classification,
  histology and application of molecular pathology}}.
\newblock {\emph{\JournalTitle{Journal of gastrointestinal oncology}}}
  \textbf{\bibinfo{volume}{3}}, \bibinfo{pages}{251} (\bibinfo{year}{2012}).

\bibitem{Lee2017}
\bibinfo{author}{Lee, J.~Y.} \emph{et~al.}
\newblock \bibinfo{journal}{\bibinfo{title}{The characteristics and prognosis
  of diffuse-type early gastric cancer diagnosed during health check-ups}}.
\newblock {\emph{\JournalTitle{Gut and Liver}}} \textbf{\bibinfo{volume}{11}},
  \bibinfo{pages}{807--812}, \doiprefix\url{10.5009/gnl17033}
  (\bibinfo{year}{2017}).

\bibitem{Chon2017}
\bibinfo{author}{Chon, H.~J.} \emph{et~al.}
\newblock \bibinfo{journal}{\bibinfo{title}{Differential prognostic
  implications of gastric signet ring cell carcinoma}}.
\newblock {\emph{\JournalTitle{Annals of Surgery}}}
  \textbf{\bibinfo{volume}{265}}, \bibinfo{pages}{946--953},
  \doiprefix\url{10.1097/sla.0000000000001793} (\bibinfo{year}{2017}).

\bibitem{tan2019efficientnet}
\bibinfo{author}{Tan, M.} \& \bibinfo{author}{Le, Q.}
\newblock \bibinfo{title}{Efficientnet: Rethinking model scaling for
  convolutional neural networks}.
\newblock In \emph{\bibinfo{booktitle}{International Conference on Machine
  Learning}}, \bibinfo{pages}{6105--6114} (\bibinfo{organization}{PMLR},
  \bibinfo{year}{2019}).

\bibitem{otsu1979threshold}
\bibinfo{author}{Otsu, N.}
\newblock \bibinfo{journal}{\bibinfo{title}{A threshold selection method from
  gray-level histograms}}.
\newblock {\emph{\JournalTitle{IEEE transactions on systems, man, and
  cybernetics}}} \textbf{\bibinfo{volume}{9}}, \bibinfo{pages}{62--66}
  (\bibinfo{year}{1979}).

\bibitem{kanavati2020weakly}
\bibinfo{author}{Kanavati, F.} \emph{et~al.}
\newblock \bibinfo{journal}{\bibinfo{title}{Weakly-supervised learning for lung
  carcinoma classification using deep learning}}.
\newblock {\emph{\JournalTitle{Scientific reports}}}
  \textbf{\bibinfo{volume}{10}}, \bibinfo{pages}{1--11} (\bibinfo{year}{2020}).

\bibitem{szegedy2016rethinking}
\bibinfo{author}{Szegedy, C.}, \bibinfo{author}{Vanhoucke, V.},
  \bibinfo{author}{Ioffe, S.}, \bibinfo{author}{Shlens, J.} \&
  \bibinfo{author}{Wojna, Z.}
\newblock \bibinfo{title}{Rethinking the inception architecture for computer
  vision}.
\newblock In \emph{\bibinfo{booktitle}{Proceedings of the IEEE conference on
  computer vision and pattern recognition}}, \bibinfo{pages}{2818--2826}
  (\bibinfo{year}{2016}).

\bibitem{kingma2014adam}
\bibinfo{author}{Kingma, D.~P.} \& \bibinfo{author}{Ba, J.}
\newblock \bibinfo{journal}{\bibinfo{title}{Adam: A method for stochastic
  optimization}}.
\newblock {\emph{\JournalTitle{arXiv preprint arXiv:1412.6980}}}
  (\bibinfo{year}{2014}).

\bibitem{tensorflow2015-whitepaper}
\bibinfo{author}{Abadi, M.} \emph{et~al.}
\newblock \bibinfo{title}{{TensorFlow}: Large-scale machine learning on
  heterogeneous systems} (\bibinfo{year}{2015}).
\newblock \bibinfo{note}{Software available from tensorflow.org}.

\bibitem{scikit-learn}
\bibinfo{author}{Pedregosa, F.} \emph{et~al.}
\newblock \bibinfo{journal}{\bibinfo{title}{Scikit-learn: Machine learning in
  {P}ython}}.
\newblock {\emph{\JournalTitle{Journal of Machine Learning Research}}}
  \textbf{\bibinfo{volume}{12}}, \bibinfo{pages}{2825--2830}
  (\bibinfo{year}{2011}).

\bibitem{Hunter:2007}
\bibinfo{author}{Hunter, J.~D.}
\newblock \bibinfo{journal}{\bibinfo{title}{Matplotlib: A 2d graphics
  environment}}.
\newblock {\emph{\JournalTitle{Computing in Science \& Engineering}}}
  \textbf{\bibinfo{volume}{9}}, \bibinfo{pages}{90--95},
  \doiprefix\url{10.1109/MCSE.2007.55} (\bibinfo{year}{2007}).

\bibitem{scikit-image}
\bibinfo{author}{van~der Walt, S.} \emph{et~al.}
\newblock \bibinfo{journal}{\bibinfo{title}{scikit-image: image processing in
  {P}ython}}.
\newblock {\emph{\JournalTitle{PeerJ}}} \textbf{\bibinfo{volume}{2}},
  \bibinfo{pages}{e453}, \doiprefix\url{10.7717/peerj.453}
  (\bibinfo{year}{2014}).

\bibitem{efron1994introduction}
\bibinfo{author}{Efron, B.} \& \bibinfo{author}{Tibshirani, R.~J.}
\newblock \emph{\bibinfo{title}{An introduction to the bootstrap}}
  (\bibinfo{publisher}{CRC press}, \bibinfo{year}{1994}).

\bibitem{goode2013openslide}
\bibinfo{author}{Goode, A.}, \bibinfo{author}{Gilbert, B.},
  \bibinfo{author}{Harkes, J.}, \bibinfo{author}{Jukic, D.} \&
  \bibinfo{author}{Satyanarayanan, M.}
\newblock \bibinfo{journal}{\bibinfo{title}{Openslide: A vendor-neutral
  software foundation for digital pathology}}.
\newblock {\emph{\JournalTitle{Journal of pathology informatics}}}
  \textbf{\bibinfo{volume}{4}} (\bibinfo{year}{2013}).

\end{thebibliography}

\section*{Acknowledgements}

We are grateful for the support provided by Professors Takayuki Shiomi \& Ichiro Mori at Department of Pathology, Faculty of Medicine, International University of Health and Welfare; Dr. Ryosuke Matsuoka at Diagnostic Pathology Center, International University of Health and Welfare, Mita Hospital; pathologists at Kamachi Group Hospitals (Fukuoka), Haradoi Hospital (Fukuoka), and Nishi-Fukuoka Hospital (Fukuoka); and Dr. Naoko Aoki (pathologist); Meng Li, Michael Rambeau, and Osamu Iizuka at Medmain Inc. We thank the pathologists who have been engaged in the annotation, reviewing cases, and pathological discussion for this study.

\section*{Author contributions statement}

F.K. and M.T. designed the studies, performed experiments, analyzed the data, and wrote the manuscript; M.T. supervised the project. All authors reviewed the manuscript.

\section*{Competing interests}

F.K. and M.T. are employees of Medmain Inc.

\end{document}